# Giant Damping-like Torque Efficiency via Synergistic Spin Hall and enhanced Orbital Hall Effects

Subhakanta Das[1], Sabpreet Bhatti[1], Ramu Maddu[1], Bilal Jamshed[1], Go Dong Wook[2] and S.N. Piramanayagam[1,*]

[1]School of Physical and Mathematical Sciences, Nanyang Technological University, 21 Nanyang Link, 637371, Singapore

[2]Peter Grünberg Institut, Forschungszentrum Jülich, Jülich, Germany.

*Corresponding author: prem@ntu.edu.sg

**Abstract**

Current-induced spin-orbit torque (SOT) has emerged as a promising method for achieving energy-efficient magnetisation switching in advanced spintronic devices. Over the past two decades, researchers have primarily focused on enhancing spin current generation through the spin Hall effect, relying predominantly on the spin degree of freedom (DoF) of the electron, while neglecting its orbital counterpart. Orbital Hall effect depends critically on the crystallinity and the interface between the orbital Hall layer and the orbital-to-spin conversion layer. However, most experimental works on orbital Hall effect relied on polycrystalline films with no special attention to improve the crystallographic texture. In this work, we have grown the Ru layer on a NiW seedlayer, which helped to improve the crystallographic texture, thereby enhancing the switching efficiency by over 44%. Such a huge increase in switching efficiency was achieved by (i) improving crystallographic texture and (ii) leveraging both spin and orbital DoFs. Our study underscores the potential for improving the spin-torque efficiency by combining interface engineering, orbital and spin Hall effects to drive next-generation spintronics.

# 1. Introduction

The advancement of emerging spintronic devices relies on the ability to control their magnetisation in a reliable and energy-efficient manner. Over the past few decades, the spin Hall effect has emerged as an efficient approach for ultra-fast magnetisation switching in heavy metal (HM)/ferromagnetic (FM) heterostructures [1–6]. In SHE, a charge current sent through the heavy metal layer is converted into a spin current through spin-orbit coupling (SOC) [6–8]. The generated spin current then flows transversely to the charge current and exerts a torque on the adjacent FM layer. The magnitude of the spin current ($I_{SH}$) strongly depends on the strength of spin-orbit coupling (SOC), which scales as $\sim Z^4$, where Z is the atomic number [9].

In a pioneering study, Liu et al. investigated current-driven magnetisation switching in Pt/Co/AlO$_x$ structures, where Pt was used as a spin Hall layer. A spin Hall angle of ~ 0.03 was reported, with a switching current density on the order of $10^{11}$ Am$^{-2}$ [10]. Concurrently, β-Ta was also explored as a spin current generator and demonstrated a reduced switching current density of the order of $10^{10}$ Am$^{-2}$, attributed to an increased SHA of 0.15 [1]. Later, two independent studies reported the highest spin Hall angles of 0.62 and 0.64 for β-W in W/CoFeB/MgO heterostructures [11,12]. However, β-W is inherently highly resistive and thus is not energy efficient despite its large SHA. Hence, sustained research efforts are required to make SOT-based devices more efficient and viable for commercial applications.

Recently, a few theoretical studies have predicted that the orbital Hall conductivity (OHC) in certain materials with weak SOC strength can exceed the spin Hall conductivity (SHC) of commonly studied high SOC materials [13–15]. Experimental validations have also confirmed that the materials, such as Zr [16,17], Ti [18–20], CuO$_x$ [21,22], Cr [23,24], Nb [20,24,25] and Ru [24–26] possess a significant OHC. Similar to SHE, the orbital Hall effect (OHE) generates an orbital current transverse to an applied electric field via non-equilibrium interband superposition of electron Bloch states with various orbital textures induced by the electric field. This process triggers the transfer of angular momentum from the lattice to the orbital part of the electron system [14,27]. However, unlike SHE, the orbital current ($I_{OH}$) generated via the OHE cannot directly interact with the local magnetisation of the magnetic material. Instead, the orbital angular momentum carried by $I_{OH}$ must be converted to spin angular momentum to exert a torque on the adjacent FM. This conversion can take place inside the FM itself or requires a heavy metal with a large SOC [20,25–28]. Therefore, a heavy metal such as Pt has been used for converting $I_{OH}$ to $I_{SH}$. For instance, R. Gupta et al. demonstrated a significant enhancement in damping-like torque efficiency by leveraging the high orbital Hall conductivity of Ru, Nb, and Cr layers in OHL/Pt/[Co/Ni]$_3$ heterostructures with perpendicular magnetisation, where Pt was used as an orbital-spin conversion layer [24].

Thus, while the discovery of OHE opens new avenues for using weak SOC materials to generate larger $I_{OH}$, the presence of strong SOC material remains essential for achieving energy-efficient magnetisation switching. Furthermore, a material with strong SOC not only facilitates the $I_{OH}$-to-$I_{SH}$ conversion but also can generate $I_{SH}$ via SHE. Therefore, by choosing the right combination of materials to form a heterostructure, it is possible to harness both OHE and SHE synergistically, thereby achieving giant damping-like torque efficiency. Nonetheless, disentangling the individual contributions of OHE and SHE remains a significant challenge.

The root cause of orbital torque generation is the coupling between the orbital part of the electron's angular momentum and the crystal lattice. The injection of orbital current from the nonmagnet to the ferromagnet critically depends on the interface crystallinity [22,27]. Therefore, it is very crucial to have the same lattice order of nonmagnetic and ferromagnetic layers. Moreover, it is critical to have a crystallographic texture in the orbital Hall layer to maximise the orbital Hall torque. However, there has not been a focus on the role of crystallographic texture in the orbital Hall effect. If we take Ru for example, when we grow it on seedlayers such as NiW alloy, Ru can be grown with a high hcp(002) texture. This is a well-exploited seedlayer in hard disk media technology to enhance Ru texture [29,30]. When most of the Ru grains are grown with their hcp(002) planes oriented parallel to the wafer plane, the orbital torque could be enhanced. However, such a study to enhance the crystalline texture has not been carried out.

This study aims to investigate the role of enhanced crystallographic texture on the enhancement in OHE. We reported a giant damping-like torque efficiency ($\xi_{DL}^{E}$) by harnessing both SHE and enhanced OHE. Further, we have carried out a qualitative analysis to disentangle the individual contributions of both effects. In a separate study, we maximised the resultant spin-torque efficiency by tuning the thickness of the spin Hall layer and observed a 1.2-fold reduction in switching current density compared to the reference sample, where only SHE was present. Furthermore, we annealed our samples to study the temperature dependence of OHE and found no appreciable change in the spin-orbit torque efficiency post-annealing. These results suggest that the studied heterostructures are thermally stable and compatible with CMOS back-end-of-line processing.

## 2. Thin-film Deposition and Characterisation

As shown in Table 1, we deposited two series of thin film samples - series M and N. Series M was engineered to understand the role of Ru hcp (002) texture on the OHE aided by a constant Pt layer in converting OHE-SHE in heterostructure stacks. The series N was aimed at maximising the resultant switching efficiency by looking at the optimum Pt layer thickness. A reference sample stack of Substrate/Ta(1)/Pt(4)/Co(1.1)/Ru(2) (thickness in nm) was also deposited for comparison.

Table 1: List of deposited samples

|    | OHE/SHE layer | FM | Capping layer |
|----|---------------|----|----|
| M1 | Ta(1)/Pt(1.5) | Co(1.1) | Pt(1.5) |
| M2 | Ta(1)/Pt(1.5) | | Pt(1.5)/Ru(4) |
| M3 | Ta(1)/Pt(1.5) | | Ru(2) |
| M4 | Ta(1)/Ru(4)/Pt(1.5) | | Ru(2) |
| M5 | Ta(1)/NiW(2)/Ru(4)/Pt(1.5) | | Ru(2) |
| N  | Ta(1)/NiW(2)/Ru(4)/Pt (1, 1.25, 1.5, 2, 2.5, 3.5, 5, 6, 8) | Co(1.1) | Ru(2) |

From symmetry considerations, sample M1 is not expected to exhibit SHE or the OHE. In contrast, sample M2 exhibits only OHE, attributed to the strong OHC of the top Ru layer. In this configuration, the top Pt layer serves as an intermediate conversion layer, transforming the $I_{OH}$-to-$I_{SH}$, which subsequently interacts with the magnetisation of the Co layer (Figure 1a). Sample M3 features only SHE, driven by the strong SHC of the bottom Pt layer. The top Ru layer in this stack is not expected to contribute significantly to the $\xi_{DL}^{E}$, as it has low SHC, and the generated $I_{OH}$ by the OHE cannot directly interact with the Co magnetisation [20,26]. On the other hand, sample M4 exhibits mainly OHE (from the bottom Ru layer), aided by the $I_{OH}$-to-$I_{SH}$ conversion by Pt. Samples M1-M4 do not have a very good crystallographic texture, as indicated by Figure 1b. Sample M5 builds upon sample M4 by incorporating a NiW seed layer, which improves the crystallographic texture of the Ru layer (Figure 1c) and thereby enhances the overall $\xi_{DL}^{E}$.

To study the effect of inserting a NiW seed layer on the crystallographic texture of Ru, we performed X-ray diffraction (XRD) measurements on our samples. We observed a minor shift in the XRD peak position of the Ru layer towards the hcp(002) phase from 2θ = 41.72° for sample M4 to 41.83° for sample M5 (Figure 1d). More importantly, the intensity of the Ru peak in sample M5 is significantly enhanced compared to that of sample M4. We also observed a significant enhancement in the intensity of the Pt fcc (111) peak at 2θ = 38.9° and Co hcp (002) peak at 2θ = 43.44° for sample M5 compared to that of sample M4 (Figure 1d). The enhancement in the peak intensity and shift in the XRD peak position of the Ru layer can be ascribed to the improvement of the crystallographic texture of the Ru layer in the presence of the NiW seed layer. NiW has been used as a seedlayer in hard disk drive technology to improve the crystallographic texture of Ru layers, to maximise the number of Ru grains with hcp(002) planes parallel to the wafer plane [29,30]. And the observed results agree with our hypothesis. Furthermore, the improved hcp phase of the Ru layer enhanced the crystal growth of the Pt and Co layer in our sample, resulting in an increment in the peak intensity [30,31].

The magnetic properties of the deposited films were studied using vibrating sample magnetometry (VSM) and magneto-optic Kerr effect (MOKE) measurements under an external out-of-plane magnetic field ($\mu_0 H_Z$) sweep. All the samples exhibit a strong perpendicular magnetic anisotropy (PMA) with square-shaped hysteresis loops (Figure 1e). Figure 1f shows their corresponding saturation magnetisation ($M_S$) values. Interestingly, we observed a lower $M_S$ for the Pt/Co/Ru samples (M3, M4 and M5) than for Pt/Co/Pt samples (M1 and M2). The reduction of $M_S$ can be attributed to the larger magnetic dead layer at Co/Ru interface relative to the Co/Pt interface, consistent with the report submitted by Zhu et al. [32]. We also observed that samples M3 and M4 showed the lowest coercivity ($H_C$), whereas sample M5 showed the highest $H_C$. The large $H_C$ of sample M5 can be attributed to the improvement of the Pt/Co interface and Pt (111) texture, as this is evident from the XRD measurements. However, the lowest $H_C$ of samples M3 and M4 (compared to M1 and M2) can be attributed to the higher intermixing of atoms at the Co/Ru interface.

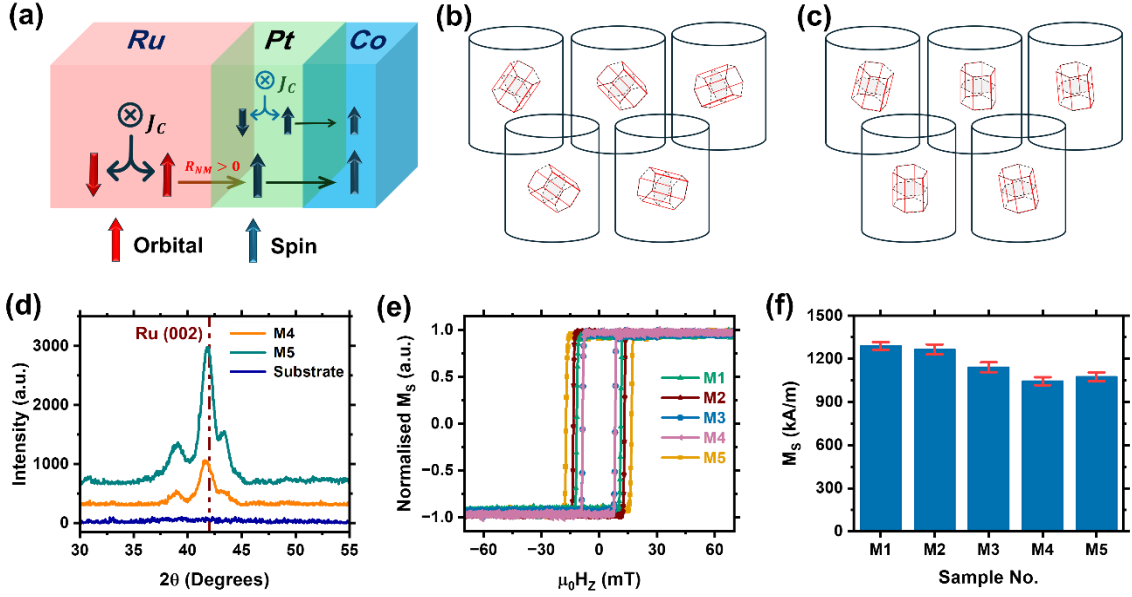

*Figure 1. (a) Schematic representations of $I_{OH}$ and $I_{SH}$ generation in Ru/Pt/Co heterostructure. $I_{OH}$ (as represented by red arrows) is generated by the strong OHE in the Ru layer. The strong SOC of the Pt layer converts the $I_{OH}$ coming from the Ru layer to $I_{SH}$ (blue arrows). Simultaneously, the Pt layer also generates an $I_{SH}$ (relatively smaller blue arrows) via the SHE. Then both the $I_{SH}$ (one generated via the SHE and the one converted from $I_{OH}$) enter and exert a spin and orbital torque on the magnetisation of the Co layer, respectively. (b) Schematic illustration of Ru grains and the orientation of hcp(002) planes of the crystal lattice (grey shaded), when Ru film is deposited in the absence of a suitable seedlayer (c) Ru grains, when the film is deposited using NiW seedlayer, which maximises the orientation of hcp(002) planes parallel to the substrate (d) XRD plots of the samples M4, M5 and bare substrate, which illustrate the effect of NiW seedlayer with a stronger hcp(002) peak intensity. (e) Normalised hysteresis loops and (f) saturation magnetisation of all the series M samples.*

### 3. Loop-shift measurements

**Sample Series-M**

We carried out current-induced loop shift measurements on Hall bar devices with lateral dimensions of 10 μm × 60 μm to calculate the $\xi_{DL}^{E}$ [33,34]. Figure 2a shows an optical microscopic image of the device with a schematic of the electrical setup. For the purpose of discussion, we present here the loop shift measurement results of samples M3, M4, and M5 only. The results for the remaining samples are provided in the supplementary information. A constant direct current (DC) pulse and an in-plane magnetic field ($\mu_0 H_X$) were applied along the longitudinal direction, and the Hall resistance was recorded as a function of $\mu_0 H_Z$ (See our previous reported work by S. Bhatti et al. for more details [34]) Figure 2b shows representative data for sample M5 at $I_{DC} = \pm 8$ mA and $\mu_0 H_X = 280$ mT. A distinct shift in the normalised anomalous Hall voltage (AHV) loop was observed, and it can be attributed to an effective out-of-plane field generated from the SOT ($\mu_0 H_{SOT}$). Figure 2c shows a plot of switching field (up-to-down, $H_{SW}^{\uparrow-\downarrow}$ and down-to-up, $H_{SW}^{\downarrow-\uparrow}$) as a function of $I_{DC}$ for $\mu_0 H_X = 280$ mT. We calculated $\mu_0 H_{SOT}$ by taking the mean value of the switching fields $[(H_{SW}^{\uparrow-\downarrow} + H_{SW}^{\downarrow-\uparrow})/2]$ for each $I_{DC}$ and performed a linear fit to calculate the slope $\mu_0 H_{SOT}/I_{DC}$ [33,34]. Figure 4d shows a plot of $\mu_0 H_{SOT}$ as a function of $I_{DC}$ for opposite polarities of an in-plane field ($\mu_0 H_X = \pm 300$ mT) and 0 mT. The opposite linear trends by changing the polarity of in-plane fields satisfy the criteria of the SOT mechanism, which originates from the opposite alignment of domain wall magnetisation in the Co layer [33].

Next, we calculated the SOT efficiency (current-induced effective field per unit electric field, $\mu_0 H_{SOT}/E$) as a function of $\mu_0 H_X$, where E is the electric field applied to the device. Figure 2e shows $\mu_0 H_{SOT}/E$ for samples M3, M4, and M5. We observed that the $\mu_0 H_{SOT}/E$ shows a 1.9-fold enhancement by utilising the OHE of the Ru layer (in M4) and a 2.7-fold enhancement by enhanced OHE with the textured Ru layer (in M5), compared to the SHE from the Pt layer (in M3). We calculated the saturation damping-like torque efficiency ($\xi_{DL}^E$) using the formula

$$\xi_{DL}^E = \frac{2e}{\hbar} M_S t_{FM} \left(\frac{\mu_0 H_{SOT}}{E}\right) \quad (1)$$

Where e is the electronic charge, $\hbar$ is the reduced Planck's constant, $\mu_0$ is the free space magnetic permeability, $M_S$ and $t_{FM}$ are the saturation magnetisation and thickness of the magnetic layer, respectively. We observed negative values for $\xi_{DL}^E$ for both samples M1 and M2. For sample M1, the net SOT is expected to be zero, as the spin current from the intrinsic SHE of the top and bottom Pt layers cancels each other out. However, the observed net $\xi_{DL}^E$ is possibly coming from the higher side-jump scattering of the electrons at the top Co/Pt interface compared to the Pt/Co interface at the bottom [7,35]. While for sample M2, the Ru layer on the top of the Co/Pt surface generates an $I_{OH}$ in the Ru layer and is converted to a spin current via the top Pt layer. As the Ru and Pt are placed above the FM layer, they have negative orbital and spin Hall conductivities, respectively. Thus, the net SOT effect in samples M1 and M2 is a negative $\xi_{DL}^E$.

The net $\xi_{DL}^E$ in samples M3 arises from the SHE of the Pt layer. Whereas, in samples M4 and M5, the $I_{OH}$ from the bottom Ru layer is converted to a positively polarised spin current via the Pt layer, supporting its intrinsic SHE. Thus, the net $\xi_{DL}^E$ in samples M3, M4 and M5 is positive. Figure 2f shows values of $\xi_{DL}^E$ of all the series M samples. We observed the $\xi_{DL}^E$ shows an enhancement of 1.7 times via OHE (comparing M2 with M1 and M4 with M3), 2.5 times via SHE and OHE (comparing M4 with M1) and 3.6 times via SHE and improved OHE (comparing M5 with M1). Further, disentangling the individual contributions of the OHE and SHE in the heterostructure, the effective $\xi_{DL}^E$ from OHE of the Ru layer is $0.39 \times 10^5$ $\Omega^{-1}$m$^{-1}$. (See supplementary for more details) Whereas that of the textured Ru layer in sample M5 is $0.81 \times 10^5$ $\Omega^{-1}$m$^{-1}$. The difference between M4 and M5 is the addition of NiW layer that enhances the Ru hcp(002) texture. A 2-fold increase of the OHE contribution to the effective $\xi_{DL}^E$, with an addition of just 2 nm NiW layer, clearly indicates the role of crystallographic texture on the OHE enhancement.

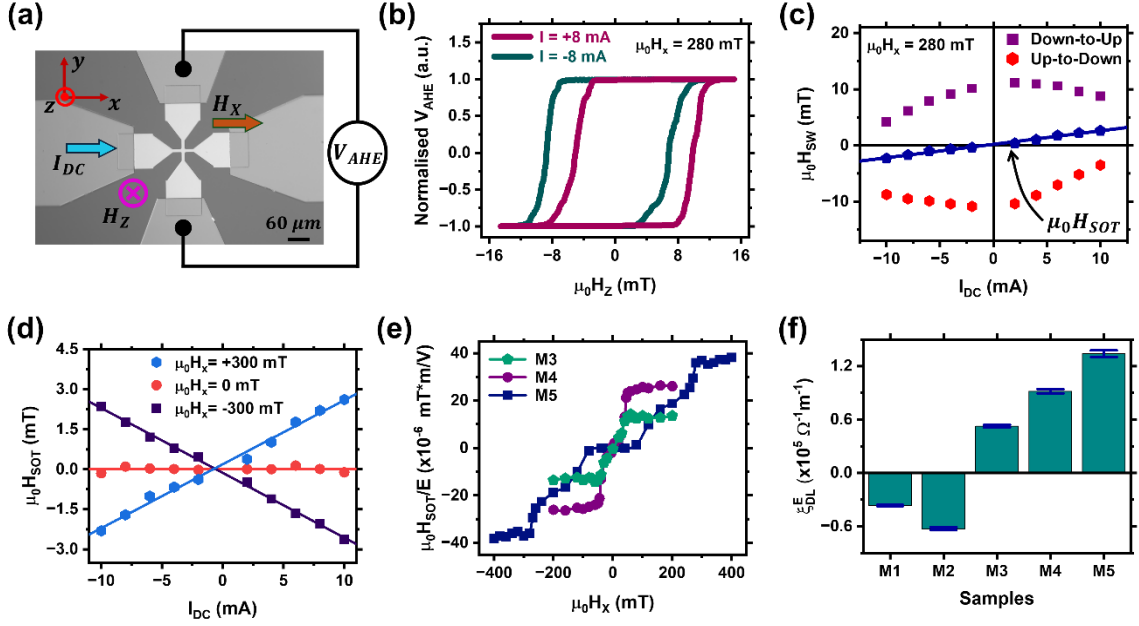

*Figure – 2. (a) Microscopic image of the Hall cross device with schematic illustrations of the measurement setup. (b) A shift in the normalised AHV loops for sample M5 for $I_{DC}$ = ±8 mA in the presence of an in-plane biasing field of $\mu_0H_X$. (c) Switching fields ($\mu_0H_{SW}$) as a function of $I_{DC}$ for switching the magnetisation from down-to-up (purple squares) and up-to-down (red hexagons) and the royal blue pentagons represent the SOT field ($\mu_0H_{SOT}$) with the straight line representing the linear fit of $\mu_0H_{SOT}$ vs $I_{DC}$ at $\mu_0H_X$ = 280 mT. (d) $\mu_0H_{SOT}$ as a function of $I_{DC}$ in the presence of $\mu_0H_X$ = -300 mT (royal blue squares), 0 mT (red circles) and +300 mT (blue hexagons) and the corresponding straight lines represent the linear fit of the data. (e) Comparison of the slope $\mu_0H_{SOT}/E$ as a function of $\mu_0H_X$ for sample M3 (green pentagons), M4 (purple hexagons) and M5 (blue squares). (f) Comparison of the $\xi_{DL}^E$ for all the series-M samples.*

**Sample Series-N**

In the earlier study, Pt thickness was kept constant, as the focus was on the role of the Ru layer and its hcp (002) texture. However, the used Pt thickness might not be the optimal thickness for the $I_{OH}$-to-$I_{SH}$ conversion. Therefore, we varied the Pt layer thickness to maximise the effective $\xi_{DL}^E$ by improving the SHE via Pt as well as the orbital to spin conversion efficiency from the Ru layer. All samples exhibited PMA with squared hysteresis loops. (See supplementary for electrical and magnetic properties characterisations) We performed similar current-induced loop shift measurements described above, to estimate $\xi_{DL}^E$.

Figure 3a shows plots of $\mu_0H_{SOT}/E$ as a function of $\mu_0H_X$, where $\mu_0H_{SOT}/E$ has the lowest value for $t_{Pt}$ = 1 nm, while the highest value is for $t_{Pt}$ = 2.5 nm among all the samples. Further, we estimated $\xi_{DL}^E$ for all the samples and plotted $\xi_{DL}^E$ as a function of $t_{Pt}$ (Figure 3d). It can be seen that $\xi_{DL}^E$ has increased monotonically with $t_{Pt}$ at the lower thickness range and reached a maximum value at $t_{Pt}$ = 2.5 nm. With further increase in $t_{Pt}$, $\xi_{DL}^E$ has saturated. $\xi_{DL}^E$ of the reference sample (horizontal line) is added for the comparison. Notably, the efficient conversion of $I_{OH}$ to $I_{SH}$ depends on an optimum thickness of the Pt layer, and the conversion efficiency reduces after a certain thickness as the spin current (that is converted from the $I_{OH}$) gets diffused by the strong SOC of the Pt layer [21,25,26]. Whereas, the spin torque by the SHE of the Pt layer may also increase with the $t_{Pt}$ at a lower thickness range and gets saturated at higher thicknesses [21,36,37]. Therefore, the enhancement of $\xi_{DL}^E$ at lower $t_{Pt}$ can be attributed to the increments of orbital-to-spin current conversion efficiency and spin current generation by the SHE of the Pt layer, reaching a maximum value for $t_{Pt}$ = 2.5 nm. The reduction of $\xi_{DL}^E$ after $t_{Pt}$ = 2.5 nm can be attributed to the decrease in orbital-to-spin conversion efficiency, as the spin current diffuses in the Pt layer. The saturation value from $t_{Pt}$ = 5 nm can be attributed to the solo contribution of the SHE of the Pt layer, as the $I_{OH}$ from the Ru layer is possibly fully diffused in the Pt layer. Thus, by optimising the Pt thickness, we obtained an enhancement of the effective $\xi_{DL}^E$ by 4 times relative to sample M3 and 1.4 times compared to the SHE of the Pt (4 nm) layer only.

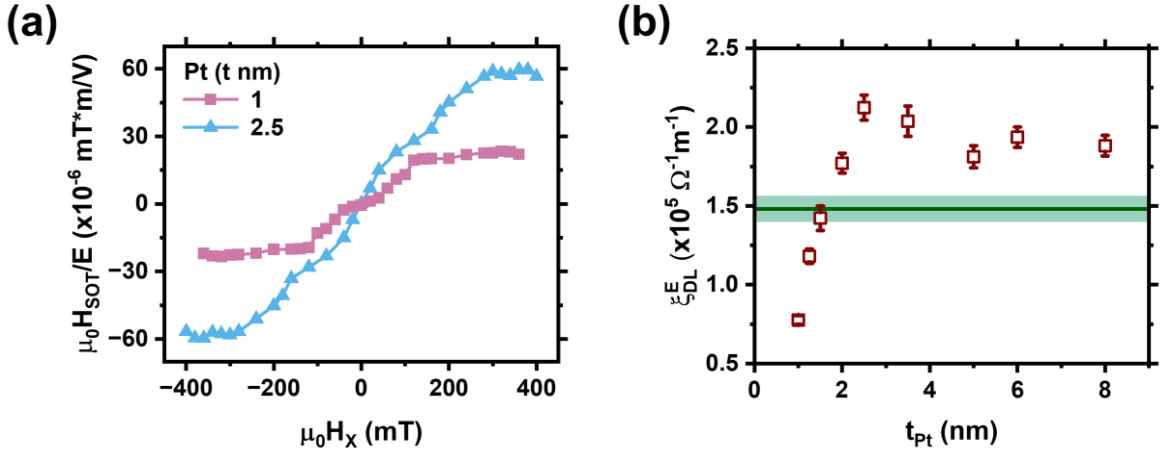

*Figure – 3. (a) Comparison of $\mu_0 H_{SOT}/E$ as a function of $\mu_0 H_X$ for samples with $t_{Pt}$ = 1 nm (reddish purple squares), $t_{Pt}$ = 2.5 nm (sky blue triangles), and $t_{Pt}$ = 8 nm (orange-red circles). (b) The trend of the $\xi_{DL}^E$ with the $t_{Pt}$ of all the series-N samples. The green line and shaded regions represent the value and error in the effective $\xi_{DL}^E$ of the reference sample, respectively.*

### 4. Current-induced Magnetisation Switching Measurements

Next, we performed current-induced magnetisation switching measurements to calculate switching current density ($J_{SW}$). We used Hall cross devices of the reference sample and a series N sample with $t_{Pt}$ = 2.5 nm (which has maximum $\xi_{DL}^E$) for the measurements. We started the measurements by saturating the magnetisation of our samples with an external out-of-plane magnetic field. Then we applied DC pulses ranging from 1 mA to 42 mA with a pulse width of 60 µs in the presence of a longitudinal $\mu_0 H_X$. After each current pulse, we applied a DC pulse of 0.15 mA with a pulse width of 0.15 s to read the anomalous Hall resistance ($R_{AHE}$) of the device. We observed squared-shaped $\Delta R_{AHE}$-J loops as shown in Figure 4a. We also observed a change in the polarity of the $\Delta R_{AHE}$-J loops by reversing the $\mu_0 H_X$ direction. Additionally, there is no magnetisation switching in the absence of $\mu_0 H_X$, suggesting that the magnetisation switching arises from the SOT of the sample [17,24,34]. We performed the first derivative on the $\Delta R_{AHE}$-J loops to calculate the switching current density ($J_{SW}$). The variation of the $J_{SW}$ as a function of $\mu_0 H_X$ for both samples is shown in Figure 4b. The experimental results showed a 1.2-fold reduction in the $J_{SW}$, from $J_{SW} = 3.01 \times 10^{11}$ Am$^{-2}$ for the reference sample to $J_{SW} = 2.51 \times 10^{11}$ Am$^{-2}$ for the Pt (2.5 nm) series N sample. We also calculated the SOT switching efficiency ($\eta$) by using equation [34,38] $\eta = [M_S t_{FM} \mu_0 (H_K - H_X)]/[2\rho d(J_{SW})^2]$. Here, $H_K$ is the anisotropy field, $\rho$ and $d$ are the resistivity and total thickness of the full stack, respectively. A higher $\eta$ indicates a more stable magnetic element, which needs lower power to switch the magnetisation. We found that the $\eta$ of the Pt (2.5 nm) series N sample is 19.6% higher relative to the reference sample. Therefore, our results suggest that choosing a suitable metal heterostructure that can utilise both the OHE and SHE together can reduce the $J_{SW}$ and enhance the power efficiency of spin devices.

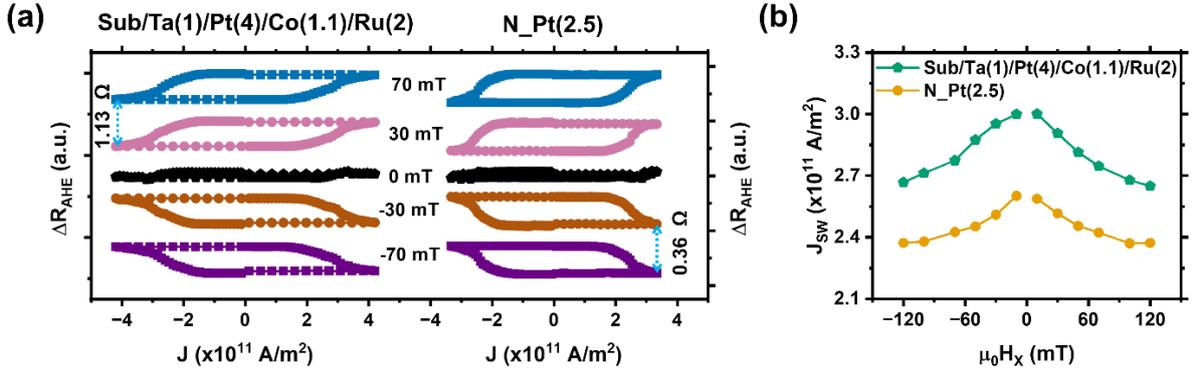

*Figure-4. (a) Current-induced magnetisation switching of the reference sample (left panel) and Pt (2.5 nm) series N sample (right panel) in the presence of $\mu_0 H_X$ ranging from –70 mT to +70 mT. (b) Variation in switching current density as a function of $\mu_0 H_X$ of the reference (blue hexagons) and Pt (2.5 nm) series N samples (red circles).*

## 5. Harmonic measurements

To validate the results obtained from the loop shift measurements, we performed AC harmonic Hall voltage measurements on a Hall cross-device with dimensions of 5 × 30 μm² to estimate $\xi_{DL}^E$. For more details on the measurement protocol, see the methods section. We performed AHE measurements to calculate anomalous Hall resistance. To estimate $\xi_{DL}^E$, we recorded the first and second harmonic responses of the device by applying AC as a function of $\mu_0 H_X$. Figures 5a, 5b and 5c show a plot of AHE, first and second harmonic responses for the sample M4, respectively. We only fitted the low-field response of first harmonic resistance data, as shown in Figure 5b, using an equation [39,40]

$$R_{XY}^{1w}(\mu_0 H_X) = R_{XY}^{1w}(0)\sqrt{1 - \left(\mu_0 H_X / \mu_0 H_K^{eff}\right)^2} \qquad (2)$$

to obtain the effective anisotropy field $\left(\mu_0 H_K^{eff}\right)$ for. Consequently, we performed second harmonic measurements as a function of $\mu_0 H_X$ and fitted the high-field data ($\mu_0 H_X$ larger than $\mu_0 H_K^{eff}$) using an equation [39–41] $R_{XY}^{2w}(\mu_0 H_X) = R_{AHE} \mu_0 H_{SOT}/2(|\mu_0 H_X| - \mu_0 H_K^{eff})$ (red curves) as shown in Figure 5c. From the fitting, we obtained the effective SOT field $\mu_0 H_{SOT}$ as

a function of alternating current applied to the device and performed linear fitting to calculate the slope $\mu_0 H_{SOT}/E$. Next, we calculated $\xi_{DL}^E$ using equation (1). We observed a similar trend of $\xi_{DL}^E$ in sample series M from both the loop shift and second harmonic measurements as shown in Figure 5d. We also performed the second Harmonic measurement on a series N sample with $t_{Pt}$ = 2.5 nm and the reference sample. From the results, we found a 1.5-fold larger $\xi_{DL}^E$ in Pt (2.5 nm) series N sample compared to the reference sample, consistent with the loop shift measurement results, as shown in the inset of Figure 5d.

Interestingly, we observe a difference in the magnitude of the $\xi_{DL}^E$ for each sample from the two measurement techniques. The difference in magnitude can be attributed to the calculation of $\mu_0 H_{SOT}$ from second harmonic measurements, where we did not separate the contribution from the planar Hall effect and the Nernst effect [40,41]. Despite the small difference, the second Harmonic measurement re-verified our claim that choosing a suitable metal heterostructure to utilise both the SHE and OHE simultaneously can further enhance the $\xi_{DL}^E$.

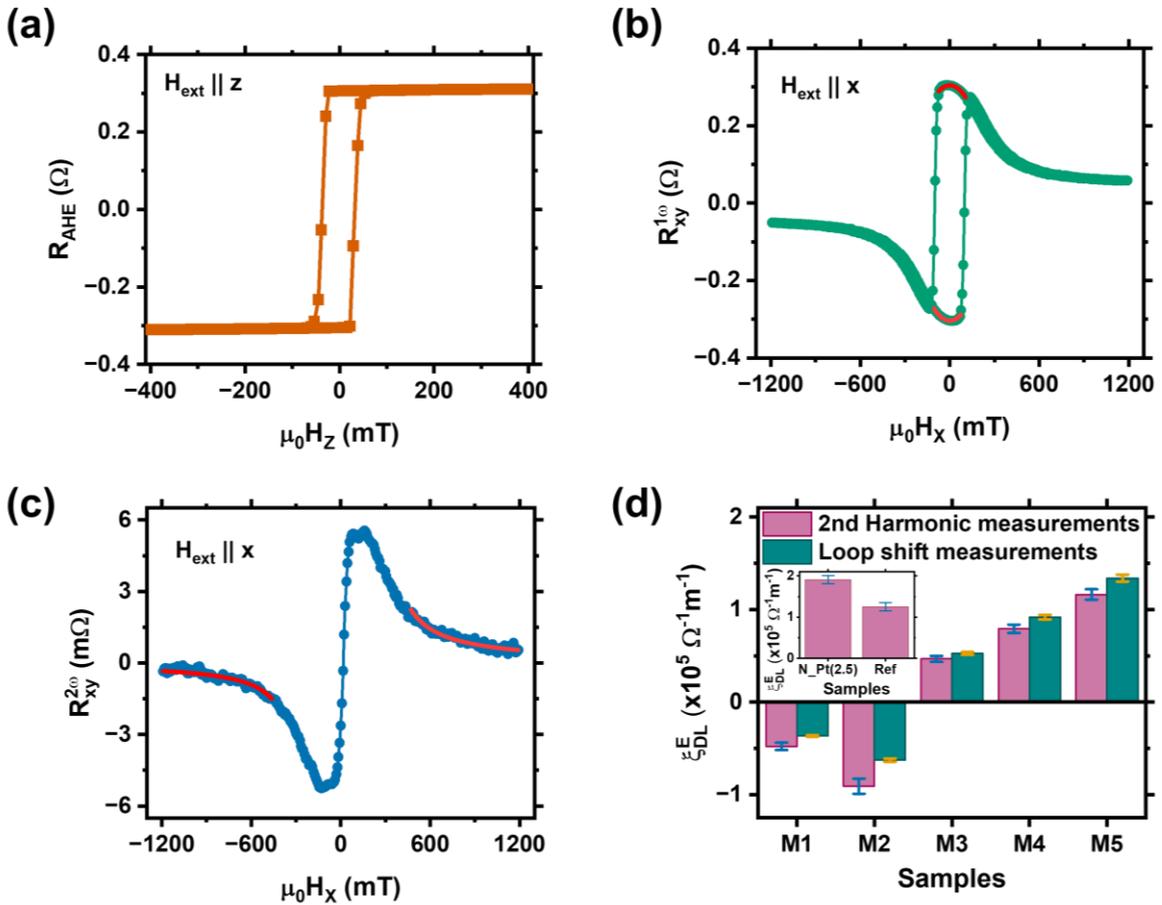

*Figure-5. AH resistance as a function the external magnetic field applied along (a) z-direction and (b) x-direction. (c) Second harmonic Hall resistance against $\mu_0 H_X$ for an AC of 8 mA for the sample M4. (d) Comparison of $\xi_{DL}^E$ for series M samples calculated from second Harmonic measurements and AHV loop-shift measurements. The inset figure represents the $\xi_{DL}^E$ of the Pt (2.5 nm) sample from the series N and the reference sample.*

6. **Effect of Annealing on the OHE**

In the production of MRAM, the magnetic tunnel junctions go through heat treatments to improve the crystallisation of CoFeB and during the packaging [42]. Therefore, it is essential to investigate the effect of temperature on $\xi_{DL}^E$. For this purpose, we annealed the series M samples in a high vacuum chamber (2 × 10$^{-8}$ Torr) at a temperature of 300 °C for one hour. After annealing, the samples were allowed to cool down to room temperature without breaking the vacuum. VSM was used to measure magnetic properties, while XRD was performed to analyse the crystal structure of Ru before and after annealing (see supplementary for details). A small shift of the Ru peak towards hcp (002) phase was observed post-annealing. Current-induced loop shift measurements were performed following the previously described procedure to quantify $\xi_{DL}^E$. A deviation in the $\mu_0 H_{SOT}/E$ was observed for the annealed sample M5 compared to its as-deposited counterpart, as shown in Figure 6a. Using equation (1), we calculate $\xi_{DL}^E$ for each sample as given in Figure 6b. Notably, a slight reduction in the $\xi_{DL}^E$ for the samples M1 and M3 was observed, where the SOT arises mainly from SHE in the Pt layer. In contrast, a significant reduction in $\xi_{DL}^E$ was observed in M2, M4 and M5, where the OHE in Ru is the dominant contribution.

Annealing of magnetic multilayers can lead to two opposing effects: (a) Improvement in the crystallinity of the Ru layer [43–45], which can enhance $\sigma_{OH}$ and increase $I_{OH}$ generation; (b) atomic intermixing at interfaces [32,46], which can reduce orbital angular momentum transfer efficiency. Therefore, the reduction of $\xi_{DL}^E$ for the annealed samples can be attributed to the interdiffusion of the metals at the Ru/Pt interface, lowering the efficient flow of $I_{OH}$ across the Ru/Pt interface. Further, the possible intermixing at the Pt/Co interface could also reduce spin transparency and thus the net flow of the $I_{SH}$ to the Co layer should be reduced. This can also be a possible reason for the net reduction of the $\xi_{DL}^E$ after sample annealing. However, the M5 sample showed comparable results for the as-deposited and annealed sample, indicating that our approach can be used for the CMOS compatible device applications.

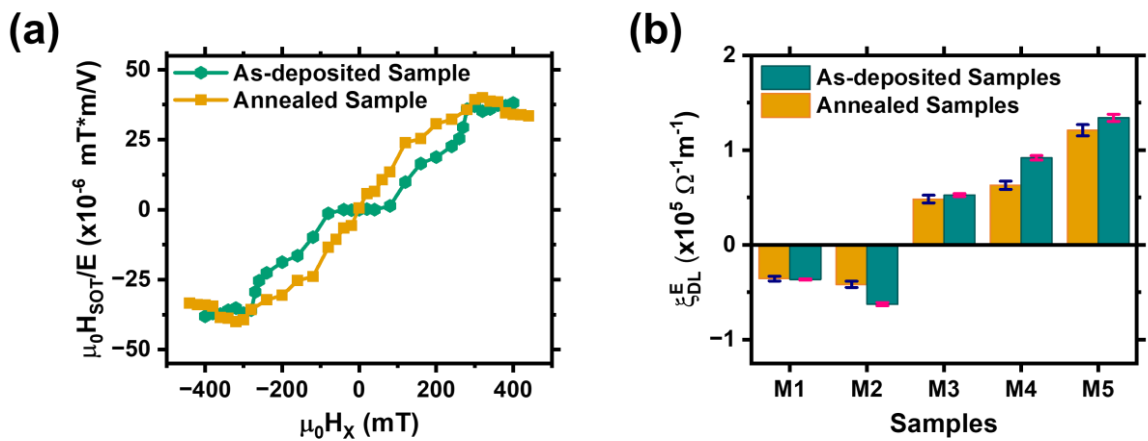

*Figure-6. (a) Comparison of $\mu_0 H_{SOT}/E$ as a function of $\mu_0 H_X$ for annealed (orange squares) and as-deposited (green hexagons) M5 sample. (b) Comparison of $\xi_{DL}^E$ for annealed and as-deposited samples of series-M.*

## 7. Conclusions

In summary, we study the influence of Ru hcp (002) texture on the enhancement of OHE arising from Ru and report a giant damping-like torque efficiency by leveraging both SHE and OHE concurrently. Further, we have carried out a quantitative analysis to disentangle the individual contributions of both effects. In a separate study, we maximised the resultant damping-like torque by tuning the thickness of the spin Hall layer and observed a 1.2-fold reduction in switching current density compared to the case where only SHE was present. Furthermore, we annealed our samples and found no appreciable change in the spin-orbit torque efficiency post-annealing. These results suggest that the studied heterostructures are thermally stable and compatible with CMOS back-end-of-line processing.

## 8. Methodology

**Thin film deposition**

We deposited all the samples mentioned in Table 1 using a DC/RF magnetron sputtering system on a Si/SiO$_2$ substrate. The substrates were sequentially cleaned by ultrasonication in acetone, isopropyl alcohol and deionised water for 2 minutes each. The base pressure of the chamber was of the order $2\times10^{-8}$ Torr. We back-sputtered the substrates for 2 minutes prior to the thin film depositions. The thin films are deposited at 50 watts of power and 3 mTorr argon pressure.

**Second Harmonic Measurements**

We used Hall cross devices of lateral dimensions 5μm×30μm for the second Harmonic measurements. We started the measurements by measuring the anomalous Hall voltage as a function of the external out-of-plane magnetic field ($\mu_0H_Z$). Then, we performed first and second harmonic measurements against a sweeping in-plane magnetic field ($\mu_0H_X$) applied along the direction of alternating current ($I_{ac}$) through the device. We calculated the $\mu_0H_{SOT}$ for each $I_{ac}$ from the fitting of the second Harmonic resistance. Then we plotted $\mu_0H_{SOT}$ as a function of E and performed linear fitting to calculate the slope $\mu_0H_{SOT}/E$. Here, E represents the electric field applied to the device to generate the flow of $I_{ac}$ through the device.